\documentclass[11pt]{article}

\input epsf.sty

\usepackage{graphicx,amsmath,amssymb}
\def\lromn#1{\uppercase\expandafter{\romannumeral#1}}

\begin{document}

\begin{center}
\begin{Large}
\textbf{
Raman stimulated neutrino pair emission
}

\vspace{1cm}

H. Hara and M. Yoshimura

Research Institute for Interdisciplinary Science,
Okayama University \\
Tsushima-naka 3-1-1 Kita-ku Okayama
700-8530 Japan

\vspace{4cm}

{\bf ABSTRACT}
\end{Large}
\end{center}

\vspace{1cm}

A new scheme using macroscopic coherence 
 is proposed from a theoretical point to experimentally determine the neutrino mass matrix, in particular
the absolute value of neutrino masses, and
the mass type, Majorana or Dirac.
The proposed process is a collective,  coherent Raman scattering followed by  neutrino-pair emission
from an excited state $|e\rangle$ of a long lifetime to a lower energy state 
 $|g\rangle$;
$\gamma_0 + | e\rangle \rightarrow \gamma + \sum_{ij} \nu_i \bar{\nu_j} + | g\rangle $
with $ \nu_i \bar{\nu_j}$ consisting of  six massive neutrino-pairs.
Calculated angular distribution has six $(ij)$
thresholds of massive neutrino-pair emission which show up as steps at different
angles in the distribution.
Angular locations of thresholds and  event rates of the angular distribution
 make it possible to experimentally determine the smallest neutrino mass to the level of less than several meV
 (accordingly all three masses using neutrino oscillation data), the mass ordering pattern, normal or inverted,
and to  distinguish whether neutrinos are of Majorana or Dirac type.
Event rates of neutrino-pair emission, when the mechanism of macroscopic coherence
amplification works,  may become large  enough for realistic experiments
by carefully selecting certain types of target atoms or ions doped in crystals.
The problem to be overcome is macro-coherently amplified quantum electrodynamic background of the process,
$\gamma_0 + | e\rangle \rightarrow \gamma +\gamma_2 + \gamma_3+ | g\rangle $,
when two extra photons, $\gamma_2\,, \gamma_3$, escape detection.
We  illustrate our idea using neutral Xe and  trivalent Ho ion   doped in dielectric crystals.

\vspace{3cm}
Keywords
 
neutrino mass matrix,
Majorana fermion,
neutrino-pair emission from atoms/ions,
coherent Raman scattering,
trivalent lanthanoid ions,
Xe metastable states


\newpage
\section
{\bf Introduction}

Remaining major  problems in neutrino physics are determination of the
absolute neutrino mass value and the nature of neutrino mass,
either of Dirac type or of Majorana type.
These problems are key important issues to clarify the origin of  baryon
asymmetry of our universe and to construct the ultimate unified theory beyond
the standard theory.
Despite of many year's experimental efforts \cite{pdg} no hint of
these issues is found so far.
It is necessary, in our opinion, to establish new experimental schemes based on targets besides  
nuclei having a few to several MeV energy release  
used in most of past experiments, since solution of these problems requires high sensitivity to the
expected, much smaller,  sub-eV neutrino mass range.

One possibility of new experimental approaches is the use of
atoms/ions or molecules
 whose energy levels can be chosen to be almost arbitrarily close to small neutrino masses 
\cite{renp overview}, \cite{my-07}.
The process is atomic de-excitation from a metastable state $ |e \rangle $ to the ground state $ |g \rangle $,
$|e \rangle \rightarrow | g\rangle + \gamma + \nu_i \bar{\nu_j} $
where $\gamma$ is detected photon accompanying invisible neutrino pair $ \nu_i \bar{\nu_j} \, ( i,j = 1,2,3) $
of mass eigenstates (anti-neutrino $\bar{\nu_i}$ is distinguishable 
from neutrino $\nu_i $ in the Dirac case, while $\bar{\nu_i} =  \nu_i$ in the Majorana case).
Necessary rate enhancement mechanism of atomic de-excitation using a coherence of
macroscopic number of atoms  (macro-coherence) has been proposed in \cite{psr th}
and its principle has been experimentally confirmed  in weak QED (Quantum ElectroDynamic) process \cite{psr exp}.
The enhancement  factor reached $10^{18}$ orders over the spontaneous emission rate.
The coherently amplified neutrino-pair emission is called RENP (Radiative Emission of Neutrino Pair),
yet to be discovered.

\begin{figure*}[htbp]
 \begin{center}
 \epsfxsize=0.5\textwidth
 \centerline{\includegraphics[width=17cm,keepaspectratio]{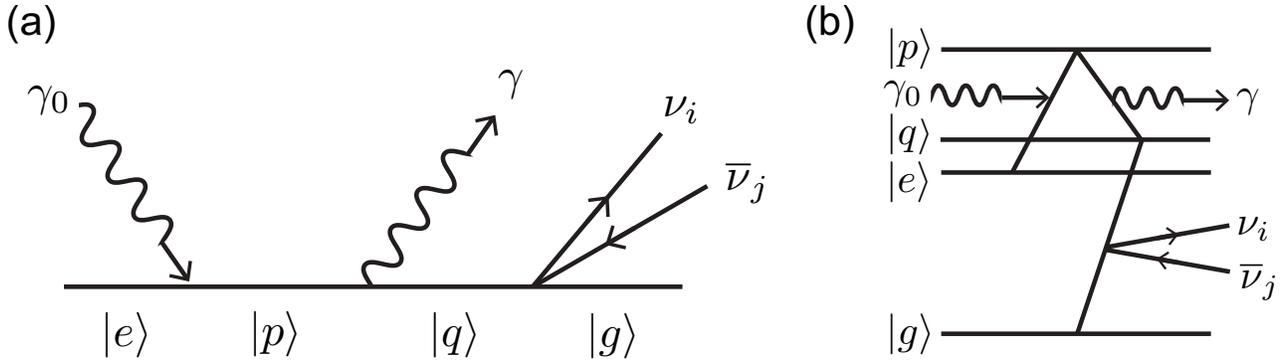}} \hspace*{\fill}
   \caption{
(a) Feynman diagram of  $ \gamma_0 + |e \rangle \rightarrow \gamma +  | g\rangle + \nu_i \bar{\nu_j} $.
 There are five more diagrams that contribute off resonances, as in eq.(\ref {atomic amplitude}).
 (b) Corresponding energy levels indicating  absorption and emission of photons and a neutrino-pair.
}
   \label {drenp feynman}
 \end{center} 
\end{figure*}

One of the problems in the original RENP scheme is
a difficulty of distinguishing a detected photon from triggered photons  (necessary to stimulate the weak process)
which happens to have the same frequency and the same emitted direction.
In the present work we study the process as depicted in Fig(\ref{drenp feynman}),
$ \gamma_0 + |e \rangle \rightarrow \gamma +  | g\rangle + \nu_i \bar{\nu_j} $,
in which the detected photon $\gamma$ has different energy and different emitted direction
from the trigger photon $\gamma_0$.
Doubly resonant neutrino-pair emission becomes possible and
massive neutrino-pair thresholds appear in the angular distribution of  detected photon $\gamma$.

The paper is organized in such a way to first present the general idea and principles of
macro-coherent neutrino-pair emission  stimulated by Raman scattering.
For brevity we call the process RANP (RAman stimulated Neutrino-Pair emission).
There are three key issues to make the RANP project of neutrino mass spectroscopy successful:
(1) mass determination and Majora/Dirac distinction \cite{my-07} is clearly possible or not,
(2) event rate is large enough or not,
(3) macro-coherently amplified QED processes that may become backgrounds are controllable or not.
Even if these issues are not ideally solved, the final question is
(4) how technological improvements may be foreseeable.
We study the general idea by using interesting
  examples of  Xe and  trivalent lanthanoid ions doped in crystals \cite{rare-earth doped crystal}, \cite{braggio},
both of which  have
large target densities typically of order $10^{20}$cm$^{-3}$ helping
for a realistic detection and have small optical  relaxation rates
for the macro-coherence amplification.
There may be other, hopefully better, candidate atoms or ions realizing the general idea, but
the atomic or ion density in a laser-excited macroscopic state must be large enough, close to a  value of
atomic density in solids for realistic detection.

We use the natural unit of $\hbar = c = 1$ throughout the present paper unless otherwise stated.

\section
{\bf Double resonance condition and Raman stimulated neutrino-pair emission}

\begin{figure*}[htbp]
 \begin{center}
 \epsfxsize=0.5\textwidth
 \centerline{\includegraphics[width=11cm,keepaspectratio]
{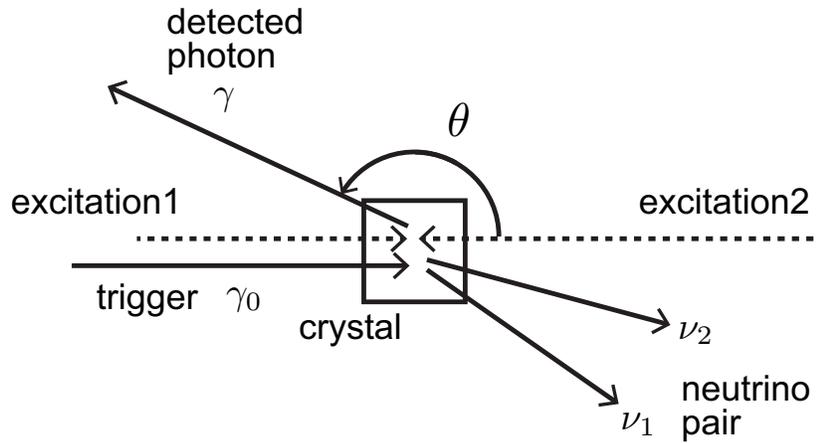}} \hspace*{\fill}
   \caption{
Schematics of experimental layout.
}
   \label {drenp schematics}
 \end{center} 
\end{figure*}

Our experimental scheme uses two counter-propagating lasers of
frequencies, $\omega_i\,,  i = 1,2$ for excitation to $| e\rangle$ from the ground state
and one trigger laser of frequency $\omega_0$ for Raman excitation, 
as illustrated in Fig(\ref {drenp schematics}).
These excitation lasers are irradiated along the same axis unit vector $\vec{e}_z $, 
hence $\omega_1 + \omega_2 =  \epsilon_{eg} $ and 
$\omega_1 - \omega_2 = r\epsilon_{eg}\,, -1 \leq r \leq 1 $.
At excitation a spatial phase 
$e^{i \vec{p}_{eg}\cdot \vec{x} }, \, \vec{p}_{eg} = r  \epsilon_{eg} \vec{e}_z$, is imprinted to target atoms,
each at position $\vec{x}$.

Suppose that a collective body of atoms, when
they have a common spatial  phase imprinted at excitation  \cite{boosted renp}, de-excite emitting plural particles,
which can be either photons or neutrino-pair.
Quantum mechanical transition amplitude, if the phase of atomic part of amplitudes,
 ${\cal A}_a = {\cal A}$, is common and uniform, is given by a formula,
\begin{eqnarray}
&&
\sum_a e^{i (  \vec{p}_{eg} + \vec{k}_0 - \vec{k} - \vec{p}_1 - \vec{p}_2 )\cdot \vec{x}_a }{\cal A}_a \simeq 
 n\, (2\pi)^3 \delta ( \vec{p}_{eg} + \vec{k}_0 - \vec{k} - \vec{p}_1 - \vec{p}_2)\, {\cal A}
\,,
\end{eqnarray}
with $n$ the assumed uniform density of excited atoms/ions.
Equality to the right hand side is valid in the continuous limit of atomic distribution.
This gives rise to the mechanism of macro-coherent amplification of rate $\propto n^2 V$
with $V$ the volume of target region.
Thus, in the macro-coherent process depicted in Fig(\ref{drenp feynman}),
 both the energy and the momentum conservation
(equivalent to the spatial phase matching condition in atomic physics terminology) hold \cite{renp overview};
\begin{eqnarray}
&&
\omega_0 + \epsilon_{eg} = \omega + E_1 + E_2
\,, \hspace{0.5cm}
\vec{k}_0 + \vec{p}_{eg} = \vec{k} + \vec{p}_1 + \vec{p}_2
\,,
\end{eqnarray}
where $E_i = \sqrt{p_i^2 + m_i^2}$ with  $m_i\,, i =1,2,3$ three neutrino masses.
From the energy and the momentum conservation one derives the kinetic region
of $(ij)$ neutrino-pair emission:
$ ( \omega_0 + \epsilon_{eg} - \omega )^ 2 -  ( \vec{k}_0 + \vec{p}_{eg} - \vec{k} )^ 2 \geq (m_i+m_j)^2$.
This may be regarded as a restriction to emitted photon energy $\omega$ and its emission angle.
At the location where the equality holds, the neutrino-pair is emitted at rest.
On the other hand, when atomic phases of ${\cal A}_a$ at sites $a$ are random in a given target volume $V$,
the rate scales with $nV$ without the momentum conservation law, which gives
much smaller  rates.

The amplitude  corresponding to Fig(\ref {drenp feynman}) 
and related five more diagrams is given by
\begin{eqnarray}
&&
{\cal A}_{\nu}(\omega) = 
- \frac{1}{ \omega_0 - \epsilon_{pe}} 
( \frac{\vec{E}_0\cdot \vec{d}_{ep} \vec{E}\cdot \vec{d}_{pq} \vec{{\cal N}}_{ij} \cdot \vec{\sigma}_{qg}  }
{  \omega - \omega_0 + \epsilon_{qe} } 
- \frac{ \vec{E}_0\cdot \vec{d}_{ep}  \vec{{\cal N}}_{ij} \cdot \vec{\sigma}_{pq} \vec{E}\cdot \vec{d}_{qg} }
{ \omega - \epsilon_{qg} } )
\nonumber \\ &&
+
\frac{1}{ \omega + \epsilon_{pe}} 
( \frac{ \vec{E}\cdot \vec{d}_{ep}\vec{E}_0\cdot \vec{d}_{pq}\vec{{\cal N}}_{ij} \cdot \vec{\sigma}_{qg}   }
{  \omega - \omega_0 + \epsilon_{qe}  } 
+ \frac{ \vec{{\cal N}}_{ij} \cdot \vec{\sigma}_{pq} \vec{E}_0\cdot \vec{d}_{qg} \vec{E}\cdot \vec{d}_{ep}}
{ \omega_0 + \epsilon_{qg}})
\nonumber \\ &&
+
\frac{1}{   \omega - \omega_0 - \epsilon_{pg} }
(\frac{ \vec{{\cal N}}_{ij} \cdot \vec{\sigma}_{ep} \vec{E}_0\cdot \vec{d}_{pq} \vec{E}\cdot \vec{d}_{qg}}
{ \omega - \epsilon_{qg}}
+ 
\frac{\vec{{\cal N}}_{ij} \cdot \vec{\sigma}_{ep}\vec{E}\cdot \vec{d}_{pq} \vec{E}_0\cdot \vec{d}_{qg} }
{ \omega_0 - \epsilon_{qg} })
\,,
\label {atomic amplitude}
\end{eqnarray} 
neglecting coupling constant factors $G_F/\sqrt{2}$.
Here $\vec{\sigma} = 2\vec{S} $ is the electron spin operator, $\vec{d} $ the electric dipole operator,
and $\vec{{\cal N}}_{ij} = \nu_i ^{\dagger}\vec{\sigma} (1-\gamma_5) \nu_j $ is 
the $(i j) $ neutrino-pair emission current arising from the spatial part of axial vector charged current 
and neutral current interaction \cite{renp overview}.
When magnetic dipole transitions are  dominant,
the electric dipole operator $\vec{d}$ in eq.(\ref{atomic amplitude}) should be replaced by the magnetic dipole operator $\vec{\mu} $.
Note that the magnetic dipole operator is odd under time reversal, while the electric dipole operator is even.
The formula, eq.(\ref{atomic amplitude}), 
is written for the level ordering $\epsilon_p > \epsilon_q  > \epsilon_e $,
but other cases of ordering may also be considered.
The energy conservation $\omega - \omega_0 =   \epsilon_{eg} - E_1 - E_2 $ can be used to rewrite 
energy denominators, for instance
$\omega - \omega_0 + \epsilon_{qe}   = - ( E_1+E_2  - \epsilon_{qg})  $.

We  find from eq.(\ref{atomic amplitude})
that double resonance occurs at $\omega_0 = \epsilon_{pe} $ and $ E_1+E_2  = \epsilon_{qg}  $
giving one diagram of Fig(\ref {drenp feynman}) dominant
(another possibility of $\omega - \omega_0 = \epsilon_{pg}  $ 
is not considered due to a difficulty of meeting the condition of McQ3 rejection later discussed).
This condition implies that $\omega = \epsilon_{pq}$.
Thus, the doubly resonant process occurs via a series of real transitions:
at first, trigger photon absorption at $|e \rangle \rightarrow | p \rangle$, followed by
a photon emission at $|p \rangle \rightarrow | q\rangle $, then by the neutrino-pair emission at
$|q \rangle \rightarrow | g\rangle $.
In the double resonance scheme
the energy denominator $\epsilon_{ab} $ should include the width factor,
$\epsilon_{ab} - i (\gamma_a + \gamma_b)/2$.

Let us take the same state for $| q\rangle = | e \rangle $.
A  feature of this scheme is that a macro-coherence exists for the last step of neutrino-pair emission,
$| q \rangle \rightarrow | g \rangle $.
One could take a view that the process is a macro-coherent neutrino-pair emission
 $ | e\rangle \rightarrow  | g \rangle +\nu\bar{\nu} $,
induced by elastic Raman scattering 
$ \gamma_0 (\vec{k}_0)  + | e \rangle \rightarrow \gamma (\vec{k})  + | e  \rangle$
of frequency $\omega_0 = \omega$, but of $\vec{k}_0 \neq  \vec{k}$.

\section
{\bf Event rate of  neutrino-pair emission and angular spectrum}

The differential spectrum rate in the double resonance scheme
 consists of a sum over contributions of massive $(ij)$ neutrino-pair production
\cite{renp overview}:
\begin{eqnarray}
&&
\frac{d^2 \Gamma_{\nu}} {d\omega d\Omega } =\frac{4 \pi^3 }{3} G_F^2
\frac{\omega^2 E_0^2 I(\omega_0) }{ \left( (\omega_0 -\epsilon_{pe})^2 + ( \gamma_{p}^2 +  \gamma_{e}^2
+ (\Delta \omega_0)^2\,)/4  \right) \left( (\omega -\epsilon_{pq})^2 + (\gamma_{e}^2+ \gamma_{q}^2)/4  \right)}
\frac{\gamma_{pe} \gamma_{pq} }{\epsilon_{pe}^3 \epsilon_{pq}^2 }
\, n^2 V  
\nonumber \\ &&
\hspace*{1cm}
\times \sum_{ij}  F_{ij} (\omega_{pe}, \cos \theta)
\Theta \left( 
{\cal M}^2 (\omega_{pe}, \theta)
- (m_i +m_j )^2\right)
\,, 
\label {diff rate0}
\\ &&
 F_{ij} = \int \frac{d^3p_1 d^3p_2} { (2\pi)^2} \delta (\epsilon_{eg} + \omega_0 - \omega - E_1-E_2)  
\delta (\vec{p}_{eg} - \vec{k} - \vec{p}_1 - \vec{p}_2)
\vec{{\cal N}_{ij}}\cdot \vec{ {\cal N}_{ij}}^{\dagger}
\,.
\label {diff rate}
\end{eqnarray}
Dependence $n^2 V $ on the excited target number density is a result of macro-coherence amplification.
The atomic part
and the neutrino-pair emission part $ F_{ij} $ are factorized in the differential rate formula, eq.(\ref{diff rate0}).
The step function 
 $\Theta\left( {\cal M}^2 (\omega_{pe}, \theta)
- (m_i +m_j )^2\right)$ determines locations of  $(ij)$ neutrino-pair production thresholds.
The squared neutrino pair current $\vec{{\cal N}_{ij}}\cdot \vec{ {\cal N}_{ij}}^{\dagger} $ 
is summed over neutrino helicities and their momenta.
We used in the formula
 experimentally measurable A-coefficients $\gamma_{ab} = (d_{ab}^2\, {\rm or} \,\mu_{ab}^2)\epsilon_{ab}^3/(3\pi)$
and the total width $\gamma_{a} = \sum_b \gamma_{ab} $ instead of dipole moments.
We denote the Raman trigger spectrum function by
$I(\omega_0) $ with width $\Delta \omega_0$ and its power $E_0^2 =
\omega_0 n_0 = \omega_0 n \eta$ where $n_0$ is the photon number density.
The dynamical factor denoted by $ \eta$ is actually time dependent,
and is calculable using the Maxwell-Bloch equation \cite{renp overview}, the coupled set of 
partial differential equations of fields and atomic density matrix elements
in the target region.
This calculation is beyond the scope of this work, and we shall assume an ideal case later on.

The quantity that appears in the formula, eq.(\ref{diff rate}), is calculated as
\begin{eqnarray}
&&
F_{ij} ( \theta)
= \frac{1}{8\pi} 
\left\{
\left(1 - \frac{ (m_i + m_j)^2}{  {\cal M}^2( \theta)} \right)
\left(1 - \frac{ (m_i - m_j)^2}{  {\cal M}^2 ( \theta)} \right)
\right\}^{1/2} 
\nonumber \\ &&
\times
\left[ \frac{1}{2} |b_{ij}|^2
\left({\cal M}^2 ( \theta) - m_i^2 - m_j^2  \right)
- \delta_M \, \Re b_{ij}^2 \,m_i m_j \right]
\,, \hspace{0.5cm}
b_{ij} = U_{ei}^*U_{ej} - \frac{1}{2} \delta_{ij}
\,.
\label {2nu integral}
\end{eqnarray}
 using methods of \cite{renp overview}.
$\delta_M =0 $ for Dirac neutrino and $=1 $ for Majorana neutrino due to identical fermion effect \cite{my-07}.
The $3 \times 3$ unitary matrix $(U_{ei} )\,, i = 1,2,3,$ refers 
to the neutrino mass mixing \cite{pdg}.
In the double resonance scheme the detected photon energy is fixed at $\omega = \epsilon_{pq} $,
and six thresholds $(ij)$ of neutrino pair production appear in the angular distribution at angles $\theta_{ij}$ of
${\cal M}^2 (\omega_{pe}, \theta_{ij}) =(m_i +m_j )^2 $.

A more practical formula for rate estimate is obtained by integrating over the detected photon energy and
convoluting with the trigger laser power.
The convolution integral over a power spectrum $ I(\omega_0)$ is
\begin{eqnarray*}
&&
\int_0^{\infty} d\omega_0 \frac{ I(\omega_0) \omega_0}
{(\omega_0 -\epsilon_{pe})^2 + ( \gamma_{p}^2 +  \gamma_{e}^2
+ (\Delta \omega_0)^2\,)/4 } \simeq 
\frac{ 2\pi I(\epsilon_{pe} ) \epsilon_{pe}}{\sqrt{ \gamma_{p}^2 +  \gamma_{e}^2 + (\Delta \omega_0)^2} }
\,,
\end{eqnarray*}
with $E_0^2 = \omega_0 \times $ the trigger photon number density,
while the integration over the detected photon energy is
\begin{eqnarray*}
&&
\int_{\epsilon_{pq}-\Delta \omega/2 } ^{\epsilon_{pq}+\Delta \omega/2 } d\omega \frac{\omega^2 }
{(\omega -\epsilon_{pq})^2 + (\gamma_{e}^2+ \gamma_{q}^2)/4  } 
\simeq \int_{-\infty } ^{\infty } d\omega \frac{\omega^2 }
{(\omega -\epsilon_{pq})^2 + (\gamma_{e}^2+ \gamma_{q}^2)/4  } 
=
\frac{ 2\pi \epsilon_{pq}^2 }{ \sqrt{ \gamma_{e}^2+ \gamma_{q}^2 }  }
\,,
\end{eqnarray*}
since the region of detected photon energy $\Delta \omega \gg \sqrt{ \gamma_{e}^2+ \gamma_{q}^2 } $.
We obtain the practical formula,
\begin{eqnarray}
&&
\frac{d \Gamma_{\nu}} { d\Omega}  \simeq
\frac{64 \pi^4}{3} 
\frac{ G_F^2   \gamma_{pe} \gamma_{pq} }
{ \sqrt{ \gamma_{e}^2+ \gamma_{q}^2 }\,\sqrt{ \gamma_{p}^2 +  \gamma_{e}^2 + (\Delta \omega_0)^2}  \,    \epsilon_{pe}^3}
n^3 V \eta  \,
4 \pi \sum_{ij} F_{ij} (\omega_{pe},  \theta)
\,.
\label {crude rate formula}
\end{eqnarray}

We shall estimate RANP rate stimulated by elastic Raman scattering.
Taking partial decay rates and total decay rates to be of the same order,
one may derive a total RANP rate scale  $\Gamma_0 $ 
by taking a typical value of  the neutrino-pair phase space integration
$\epsilon_{eg}^2/24 $ from eq.(\ref{2nu integral}) approximated in the massless neutrino limit,
\begin{eqnarray}
&&
\frac{\Gamma_0}{4\pi}  = 
\frac{32 \pi^4}{3\sqrt{2} } 
\frac{ \gamma_{pe}^2 } { \sqrt{ \gamma_{p}^2 +  \gamma_{e}^2 + (\Delta \omega_0)^2}      } \frac{1}{ \gamma_e}
 \, \frac{G_F^2  \epsilon_{eg}^2 }{ 24 \epsilon_{pe}^3}\, n^3 V \eta
\,.
\label {ranp rate unit}
\end{eqnarray}

The rate formula calculated this way contains, besides detector and
laser related quantities, four important factors, and
these  appear in the rate  as

rate = Raman scattering rate ($ \gamma_{pe}^2/(\sqrt{ \gamma_{p}^2 +  \gamma_{e}^2 + (\Delta \omega_0)^2}  )\,) 
\times $ lifetime of $|e \rangle ( = | q\rangle )$ state
($1/\gamma_e$) 
$\times $ neutrino-pair emission rate ($G_F^2\epsilon_{eg}^2/ \epsilon_{pe}^3) \times$ coherence amplification factor
($n^3 V \eta $),

in the neutrino-pair emission stimulated by elastic Raman scattering.
From eq.(\ref{ranp rate unit}) it becomes  very important for target selection 
how large a dimensionless combination of decay rate, lifetime and energy differences,
$\gamma_{pe}^2 \epsilon_{eg}^2/ (\gamma_e \epsilon_{pe}^3 )   $ is.

\section
{\bf Amplified QED backgrounds}

The macro-coherent amplification  necessary for  RANP  rate enhancement
may also amplify QED processes which may give rise to serious backgrounds.
These amplified QED processes are termed as McQn (macro-coherent QED n-th order photon emission) \cite{mcqn}.
We shall first consider how to get rid of MacQn (n=2,3) backgrounds.

The  quantity 
${\cal M}^2 = (\omega_0 + \epsilon_{eg} - \omega )^2 - ( \vec{k}_0 + \vec{p}_{eg} - \vec{k})^2 $ is
equal to $(E_1 + E_2)^2 - (\vec{p}_1 + \vec{p}_2)^2$ using variables of unseen $(\nu_1, \nu_2)$ and
is important for discussion of backgrounds. In terms of Raman scattering variables it  reads
as
\begin{eqnarray}
&&
{\cal M}^2(\theta)  = 
(1-r^2) \epsilon_{eg}^2 - 2 (1-r) \epsilon_{eg} (\omega - \omega_0)
- 4\omega ( r  \epsilon_{eg} + \omega_0) \sin^2 \frac{\theta}{2}
\,,
\label {mcq3 squared mass}
\end{eqnarray}
where $\theta$ is the $\gamma$ emission angle measured from the excitation axis.
When a photon of 4-momentum $k = (\omega, \vec{k})$ is emitted instead of the neutrino-pair,
this quantity vanishes at some angle satisfying ${\cal M}^2 (\theta) (= \omega^2 - \vec{k}^2) = 0 $.
In this case a serious
amplified McQ3 background exists.
On the other hand, if this quantity is  arranged to be positive at all angles and is taken close to
neutrino mass thresholds, $(m_i+m_j)^2 $,  neutrino-pair emission occurs without the
McQ3 background. 
With a proper choice of the imprinted phase  $r$ and trigger frequency $ \omega_0$, one
may readily work out the parameter region that excludes  QED backgrounds of
McQ3. The other background, McQ2 (Paired Super-Radiance) 
$  |e\rangle \rightarrow |g \rangle + \gamma_0 +\gamma $, is rejected  unless
$r = - 1 + 2\omega_0/\epsilon_{eg}$, which we shall assume to be valid in the following.

In the case of elastic RANP ($\omega = \omega_0$) 
the first angular threshold rise occurs at pair production of smallest neutrino mass $m_1$
at an angle,
\begin{eqnarray}
&&
 \sin^2 \frac{\theta}{2} = \frac{ (1-r^2) \epsilon_{eg}^2 - 4 m_1^2}{4 \omega_0 (r \epsilon_{eg} + \omega_0)  }
\,.
\label {th angle}
\end{eqnarray}
The formula may be used in other situations.
A light hypothetical particle X such as axion and hidden photon \cite{hidden photon}
can be searched as an angular peak given by the angle $\theta_X$ obtained
by replacing $4 m_1^2$ in eq.(\ref {th angle}) by the squared X-mass $m_X^2$

Without McQ2 and McQ3 events the largest amplified QED background arises from McQ4,
$\gamma_0 + | e\rangle \rightarrow \gamma +\gamma_2 + \gamma_3+ | g\rangle $,
obtained  by replacing the neutrino-pair $\nu_i \bar{\nu_j} $
in $|q \rangle \rightarrow |g\rangle$ by two photons, $\gamma_2 \gamma_3$.
It is difficult to kinetically reject McQ4 events when two extra photons
escape detection, since the two-photon system has its squared pair-mass  
coincident to RANP events.
The crucial question is how  big the McQ4 background rate is.
Disregarding the macro-coherent amplification factor common to RANP and McQ4 event rates,
\begin{eqnarray*}
&&
\frac{8 \pi^3 }{3}
\frac{\omega^2 E_0^2  I(\omega_0)}{ \left( (\omega_0 -\epsilon_{pe})^2 + ( \gamma_{p}^2 +  \gamma_{e}^2
+ (\Delta \omega_0)^2\,)/4  \right) \left( (\omega -\epsilon_{pq})^2 + (\gamma_{e}^2+ \gamma_{q}^2)/4  \right)}
\frac{\gamma_{pe} \gamma_{pq} }{\epsilon_{pe}^3 \epsilon_{pq}^2 }
\, n^2 V  
\,,
\end{eqnarray*}
one should compare two quantities,  the RANP rate function
$I _{2\nu} = G_F^2 \sum_{ij} F_{ij}/2$ already given and
McQ4 rate function, which has two possibilities, either  E1 (electric dipole) $\times $ M1 (magnetic dipole)
or E1 $\times $ E1.
Macro-coherent E1 $\times $ E1 rate function (one electric dipole $d$ replaced by the magnetic dipole $\mu$ for  E1 $\times $ M1)
$I _{2\gamma} $ to be compared with the factor of RANP function $I _{2\nu}$ is
\begin{eqnarray}
&&
I_{2\gamma} =
\int \frac{d^3 k_2 d^3 k_3}{(2\pi)^2}\sum_n 
\frac{d_{ne}^2 d_{ng}^2}{ (\omega_2 + \epsilon_{ne} )^2} \frac{\omega_2 \omega_3}{6^2}
\delta (\omega_2 + \omega_3 - \epsilon_{eg}) \delta (\vec{k}_2 + \vec{k}_3 - \vec{P})
\,,
\end{eqnarray}
with $\vec{P} = \vec{p}_{eg} + \vec{k}_0 - \vec{k} $.
The result of calculation is given by
\begin{eqnarray}
&&
I_{2\gamma}({\cal M}^2)  = \frac{\pi}{8}  \sum_n \frac{\gamma_{ne} \gamma_{ng} }{\epsilon_{ne}^3 \epsilon_{ng}^3}
\frac{1 }{\sqrt{\epsilon_{eg}^2 -  {\cal M}^2} } \int_{\omega_-}^{\omega_+} d\omega
\frac{\omega^2 ( \epsilon_{eg} - \omega)^2  }{(\omega + \epsilon_{ne} )^2 }
\,, \hspace{0.5cm}
\omega_{\pm} = \frac{1}{2} \left(  \epsilon_{eg} \pm \sqrt{\epsilon_{eg}^2 -  {\cal M}^2 }
\right) 
\,,
\label {mcq4 related formula}
\end{eqnarray}
where the relation $\vec{P}^2=  \epsilon_{eg}^2 -  {\cal M}^2$ was used.
Upper energy states $| n \rangle$ in eq.(\ref{mcq4 related formula})
must be connected to lower energy states by
the combination of E1 $\times $ E1 for Xe case and E1 $\times $ M1 for
trivalent lanthanoid ion, which restricts states of large contributions \cite{backgrounds}.

\section
{\bf Numerical estimate of absolute RANP angular distributions and background rates for Xe and Ho$^{3+}$ doped in crystals}

\begin{figure*}[htbp]
 \begin{center}
 \centerline{\includegraphics[width=14cm,keepaspectratio]{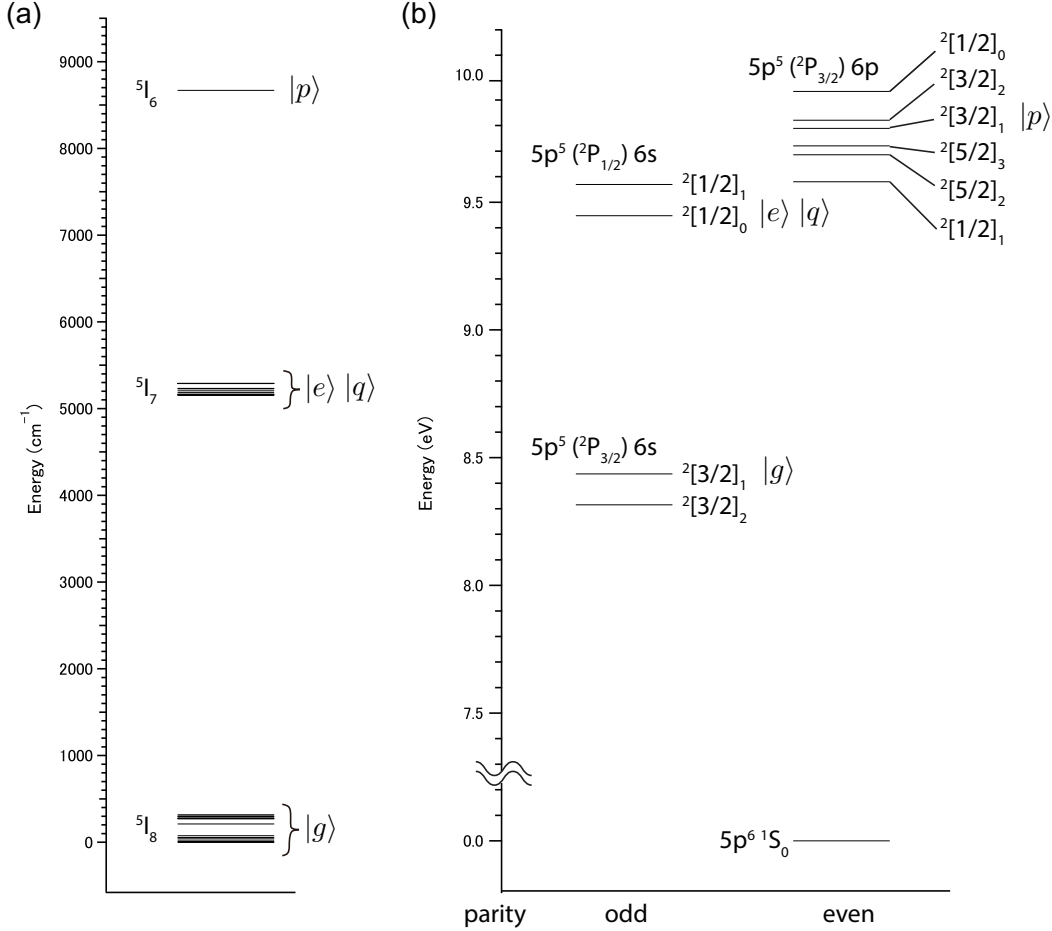}} \hspace*{\fill}
   \caption{
Relevant energy levels of trivalent Ho ion in YLF and neutral Xe.
(a) Ho in the unit of cm$^{-1}$ ($10^4{\rm cm}^{-1} = 1.24 {\rm eV} $ in the natural unit). 
There are a number of split Stark states under crystal field that contribute as states, $|q \rangle$ and $ | g \rangle $. (b) Xe in eV unit.
}
   \label {xe-ho levels}
 \end{center} 
\end{figure*}

In this section we apply theoretical formulas given in the preceding sections to real target atoms/ions.
An exhaustive study of RANP and background rates  is beyond the scope of this work.
We shall restrict to two interesting cases of a trivalent lanthanoid ion and Xe atom
whose relevant energy levels are depicted in Fig(\ref{xe-ho levels}).
As shown there, there are a number of Stark states (degenerate states lifted by crystal field) that may contribute to RANP process:
11 theoretically expected and 9 experimentally detected levels for $|p \rangle = ^5$I$_7$, and
13 theoretically expected and 9 experimentally detected levels for $|g \rangle = ^5$I$_8$ \cite{ed in lanthanoid}.
Trivalent lanthanoid ions have a number of $4f$ electrons shielded by outer $6s$ electrons,
giving rise to sharp line widths, when they are doped in dielectric crystals \cite{rare-earth doped crystal}.
Xe has two metastable excited states, $2^-$ (lifetime $\sim 43$ sec)
and $0^-$ (lifetime $\sim 0.13$ sec), suitable for RENP \cite{renp overview} and RANP.
Both targets can be prepared to have large number densities.

\subsection
{\bf Lanthanoid case: example of  Ho$^{3+}$ doped in YLF}

We first comment on the important quantum number of state classification in solids.
Without a magnetic field application (and even in the presence of an internal magnetic field
of nucleus) time-reversal symmetry holds, but
parity may be violated in the presence of  the crystal field.
Unlike the  state classification in terms of  parity in the free space (vacuum) one should
use time-reversal quantum number,
even or odd, or T quantum number in short.
Hence optical transitions between two Stark states of definite T quantum numbers, 
either inter- or intra-J manifolds,
should be classified according to relative T quantum numbers,  even or odd.
Following this classification T-odd single photon emission  goes via M1,
while T-even emission goes via E1 or  E2 (electric quadrupole).
Transitions among states made of $4f$ electrons in the free space
are  mainly M1, but parity violating effects caused by crystal field
make E1 often dominant in crystals, as shown in \cite{ed in lanthanoid},  \cite{judd-ofelt}.

The last step in the resonant path, $|q \rangle \rightarrow | g \rangle $, must
be M1 due to the nature of neutrino-pair emission operator, the spin of electron $\vec{S}_e $.
In the trivalent Ho ion transition paths of M1 nature are limited: \\
$^5{\rm I}_7 \rightarrow  ^5\!{\rm I}_8\,, ^5{\rm I}_6 \rightarrow  ^5\!{\rm I}_7 \,, ^5{\rm I}_5 \rightarrow  ^5\!{\rm I}_6\,,
^5{\rm I}_4 \rightarrow  ^5\!{\rm I}_5\,,    ^5{\rm F}_4 \rightarrow  ^5\!{\rm F}_5$ 
from the list of \cite{ed in lanthanoid}.
Other steps, $| e \rangle \rightarrow | p \rangle$ and $|p \rangle \rightarrow | q \rangle$,
should be chosen from large listed  A-coefficients, often from E1 transitions.

We first discuss McQ4 background whose amplitude is obtained by replacing neutrino-pair emission
at $|q \rangle \rightarrow | g \rangle$ by two-photon emission.
T-odd two-photon emission occurs dominantly via M1 $\times $ E1.
Parity violating effect due to crystal field and consequent weak E1 decay rate calculation 
was formulated in \cite{judd-ofelt}, and decay rates among J-manifolds has been 
given  in \cite{ed in lanthanoid} where we can find almost all data we need for our calculation.
In trivalent Ho ion there are not many common levels $|n \rangle$ that have E1 and M1 coupling to
 $|q \rangle \,, | g \rangle$.
From the point of background rejection it is desirable to search for  $|q \rangle \,, | g \rangle$
which have no sizable M1 $\times $ E1.
Indeed, there are a few candidates of this property.
Another consideration we have to focus on is to choose the rate factor,
$\gamma_{pe}\gamma_{pq}/\sqrt{\gamma_e^2 + \gamma_q^2}$, as large as possible.

A choice of RANP path considering McQ3 rejection and large RANP rate is for
neutrino-pair emission stimulated by elastic Raman scattering,
\begin{eqnarray}
&&
| e \rangle =^5{\rm I}_7\,   5152 {\rm cm}^{-1} \rightarrow 
 |p\rangle =^5\!{\rm I}_6\, 8669 {\rm cm}^{-1}
\rightarrow |q \rangle = | e\rangle  \rightarrow 
 |g\rangle = ^5\!{\rm I}_8\, 0.1 {\rm cm}^{-1}
\,.
\label {ho path}
\end{eqnarray}
The useful relation in the natural unit is $10^4{\rm cm}^{-1} = 1.24 {\rm eV} $.
Energy values are taken from \cite{ed in lanthanoid}.
Lowest energy Stark levels of J-manifolds are chosen to make effects of phonon emission minimal.
We are not informed of T quantum numbers, hence if some of them do not match T-odd
or T-even rule,
different Stark levels in the vicinity must be changed to.
To the calculation accuracy of \cite{ed in lanthanoid}  M1 $\times $ E1 two-photon
emission in $|e\rangle \rightarrow |g \rangle $ transition is forbidden,
hence there is no McQ4 background to this accuracy.

\begin{figure*}[htbp]
 \begin{center}
 \centerline{\includegraphics{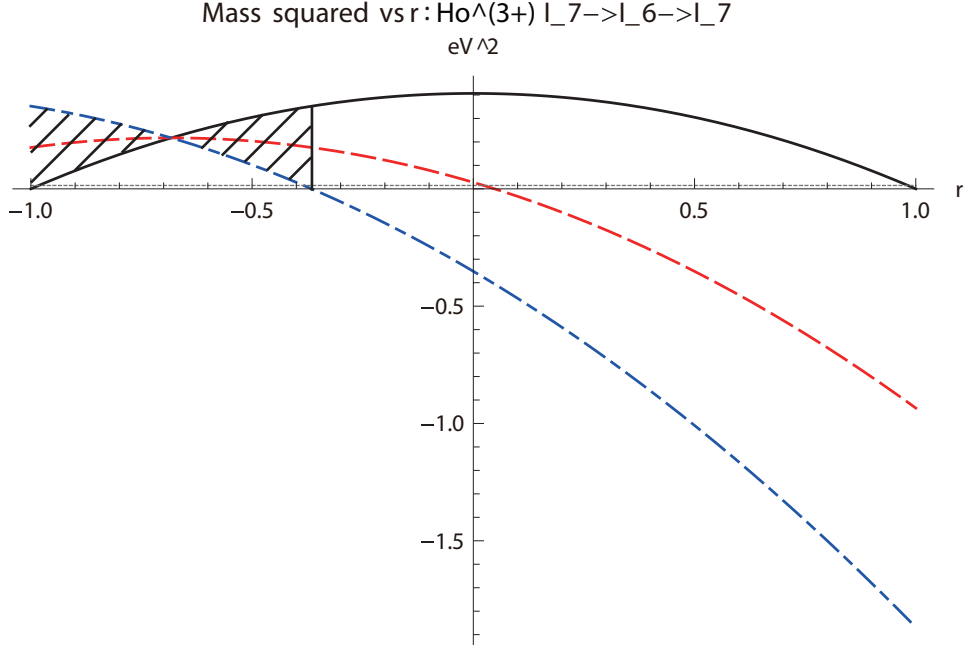}} \hspace*{\fill}
   \caption{
Mass squared ${\cal M}^2(\theta; r)$ vs $r$ for three angles:
$\theta = 0$ in solid black, $\theta = \pi/2$ in dashed red, and $\theta = \pi$ in dash-dotted blue.
The line ${\cal M}^2(\theta; r) = 4 (60 {\rm meV})^2$ (expected (33) mass square ${\cal M}^2$ assuming
the smallest neutrino mass $\ll$ 10 meV ) is shown in dotted black as a guide.
It is concluded that no McQ3 is expected for $r \leq -0.36532$ as shown in the shaded region
where ${\cal M}^2(\theta; r) >0 $ for all angles.
At $r = - \epsilon_{pe}/\epsilon_{eg} $ there is no dependence of rates on emission angle.
}
   \label {sq mass 1}
 \end{center} 
\end{figure*}

\begin{figure*}[htbp]
 \begin{center}
 \centerline{\includegraphics{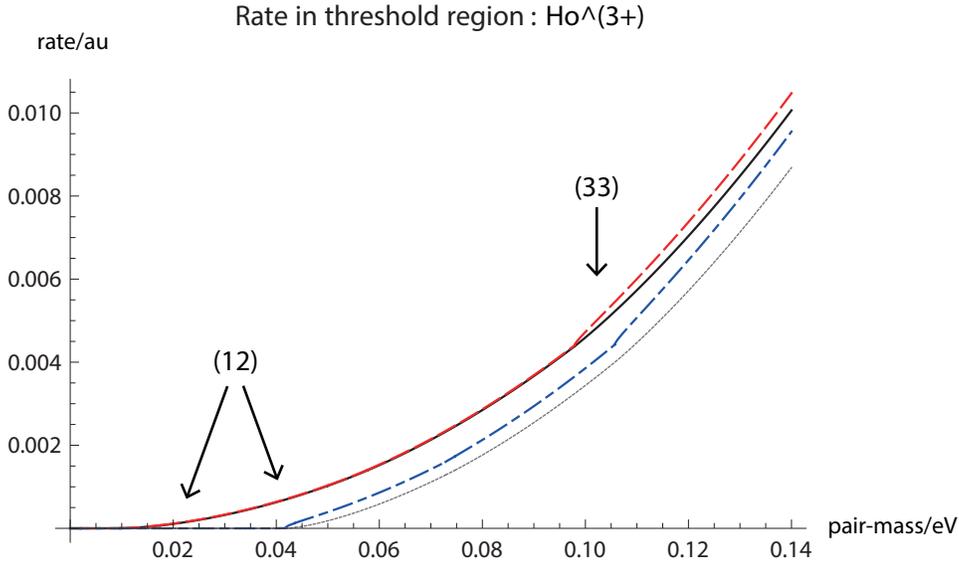}} \hspace*{\fill}
   \caption{
Rate vs neutrino-pair mass $\sqrt{ {\cal M}^2(\theta; r)} $ with $r = -0.36532$ of trivalent Ho.
Cases of NO Majorana of smallest neutrino mass 1 meV in solid black,
NO Majorana of 50 meV in dotted black,
NO Dirac 1 meV in dashed red, and NO Dirac 50 meV in dash-dotted blue.
Two arrows indicate the largest threshold rises of (12) and (33) pair-production.
}
   \label {sq mass distribution 1}
 \end{center} 
\end{figure*}

\begin{figure*}[htbp]
 \begin{center}
 \centerline{\includegraphics{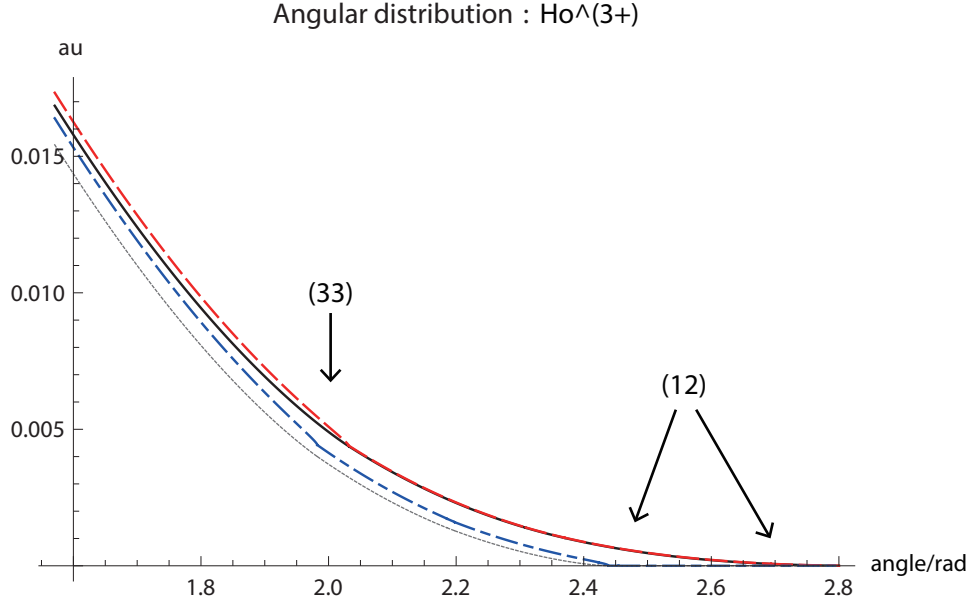}} \hspace*{\fill}
   \caption{
RANP angular distribution with $r = -0.36532$ of trivalent Ho.
Cases of NO Majorana of smallest neutrino mass 1 meV in solid black,
NO Majorana of 50 meV in dotted black,
NO Dirac 1 meV in dashed red, and NO Dirac 50 meV in dash-dotted blue.
The absolute rate is obtained by multiplying a factor $5.1  \times 10^{-6} $sec$^{-1}$ assuming
$n^3 V \eta = (10^{17} )^3 $cm$^{-6}$.
This rate unit is applied to Fig(\ref{sq mass distribution 1}) and Fig(\ref{ranp angular distribution nio1}) $\sim$ 
 Fig(\ref{ranp angular distribution from different path 1}) as well.
}
   \label {ranp angular distribution 1}
 \end{center} 
\end{figure*}

\begin{figure*}[htbp]
 \begin{center}
 \centerline{\includegraphics{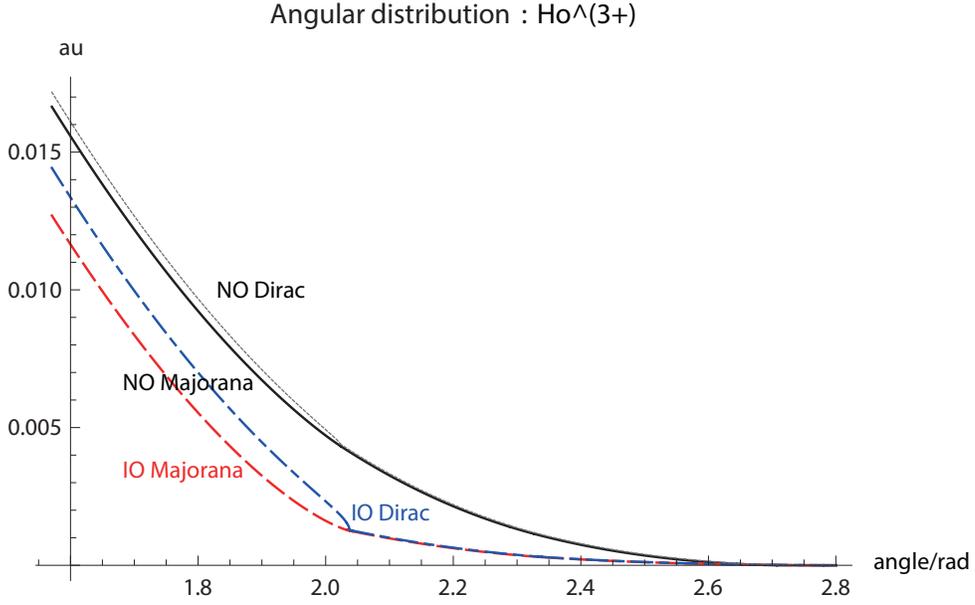}} \hspace*{\fill}
   \caption{
RANP angular distribution with $r = -0.36532$ of trivalent Ho.
Cases of NO Majorana of smallest neutrino mass 5 meV in solid black,
NO Dirac of 5 meV in dotted black,
IO Majorana 5 meV in dashed red, and IO Dirac 5 meV in dash-dotted blue.
}
   \label {ranp angular distribution nio1}
 \end{center} 
\end{figure*}

\begin{figure*}[htbp]
 \begin{center}
 \centerline{\includegraphics{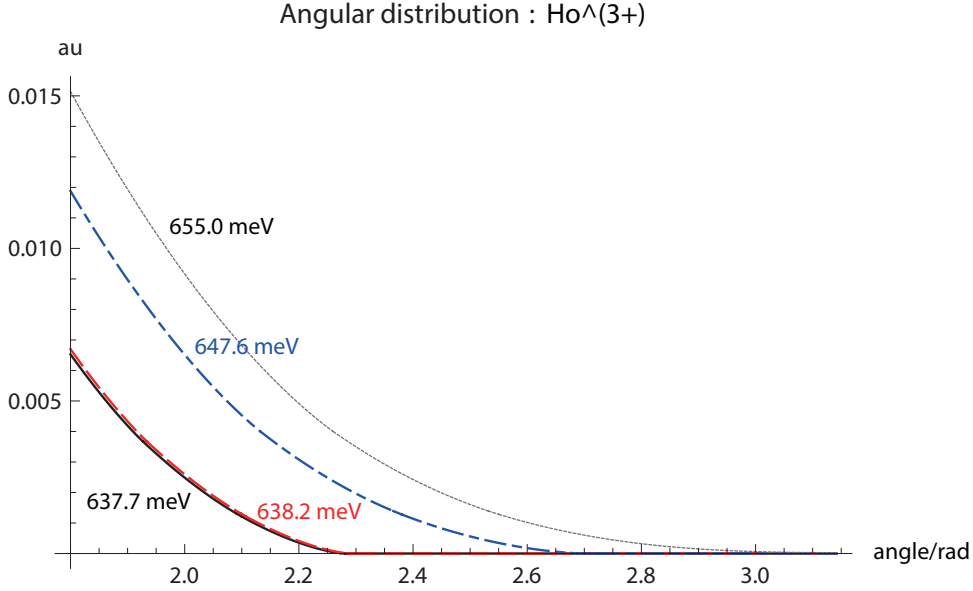}} \hspace*{\fill}
   \caption{
RANP angular distribution with $r = -0.36532$ of trivalent Ho 
from different inelastic Raman paths of $|q \rangle =  ^5$I$_7, 655.0\,$meV,  647.6\,meV, 
638.2 meV, and 637.7 meV in the case of
 NO Majorana of smallest neutrino mass 30 meV.
}
   \label {ranp angular distribution from different path 1}
 \end{center} 
\end{figure*}

We first show the angular distribution given by $\sum_{ij}  F_{ij} (\omega_{pe}, \cos \theta) $ of eq.(\ref{2nu integral})
with ${\cal M}^2$ of eq.(\ref{mcq3 squared mass}).
The angular spectrum is sensitive to an adopted value of the imprinted phase factor $r$.
We investigated this dependence for a few trivalent lanthanoid ions
doped in crystals  by calculating the squared  mass ${\cal M}^2 (\theta; r)$
and searched for the parameter $r$ to optimize the  shape of angular spectrum clearly showing neutrino-pair threshold rises.
The search is illustrated in 
Fig(\ref{sq mass 1}), which gives an optimal value, $ r=  -0.36532$.
We show for this $r$ choice
the squared mass ${\cal M}^2$ distribution in Fig(\ref {sq mass distribution 1}) and the angular distribution
in Fig(\ref{ranp angular distribution 1}) $\sim $  Fig(\ref{ranp angular distribution from different path 1}).
The two largest threshold rises appear at the pairs, (12) and (33), where
$|b_{12}|^2 = 0.405\,, |b_{33}|^2 = 0.227 $, making up most of the weight sum $\sum_{ij} | b_{ij}|^2 = 3/4$
\cite{nu parameters}.
The smallest neutrino mass of order 1 meV is best determined 
by measurements around (12) threshold, as is made evident in Fig(\ref{sq mass distribution 1})
and Fig(\ref{ranp angular distribution 1}), while
the Majorana/Dirac distinction is better studied after (33) threshold opens.
The mass ordering pattern \cite{pdg}, normal  ordering (NO) vs inverted ordering(IO) distinction,
 is relatively easy as seen in Fig(\ref {ranp angular distribution nio1}).

The unique feature of RENP and RANP experiments is that there is a sensitivity to
determine CP violation phases intrinsic to Majorana neutrinos.
We have not, however, studied this sensitivity in the present work.

With the $r$ choice of these figures,  frequencies of two excitation lasers are
$\omega_1 = 202. 366$ meV $, \omega_2 =435.329 \,$ meV, while the Raman trigger 
and detected photon have energies,  $\omega_0 = \omega = 435.33  \, $meV.
In Fig(\ref{ranp angular distribution from different path 1}) we show contributions from
inelastic RANP paths arising from different, wide spread, Stark states
for $|q \rangle$ which should be separately detectable with a high resolution
of detected photon energy.
These inelastic Raman paths (contributions beside the one in solid black of Fig(\ref{ranp angular distribution from different path 1})\,)
give rise to pair production at finite neutrino velocities.
These contributions refer to production far away from thresholds, hence
they are not sensitive to neutrino mass determination.  But
they are important to identify the process of macro-coherent neutrino-pair emission in atoms/ions.
Other paths from
Stark states in manifolds, $^5$I$_7$ and $^5$I$_8$, of the same T quantum numbers
should equally contribute to RANP photon angular distributions.

The RANP rate scale $\Gamma_0$ can be calculated using the formula, eq.(\ref{ranp rate unit}), with \\
$\epsilon_{eg} =  631\,  {\rm meV}\,, \epsilon_{pe} =431\, {\rm meV}\,, \gamma_{pe} = 14.2\, {\rm sec}^{-1} \,,
\gamma_{e} = 69.9\, {\rm sec}^{-1}$ (the lifetime $1/\gamma_{e} $ 
including non-radiative contributions at 10 K) for the relevant path.
The result is 
\begin{eqnarray}
&&
\Gamma_0({\rm Ho} )  = 0.85 \times 10^{-7}   \,{\rm sec}^{-1}
\frac{2\pi \, 100 {\rm Hz} }{\Delta \omega_0 } \frac{ n^3 V \eta }{ (10^{17})^3\, {\rm cm}^{-6}}
\,.
\end{eqnarray}
We assumed that laser related factors can give the combination $ n^3 V \eta$ of order $ (10^{17})^3\, {\rm cm}^{-6}$,
having in mind 10 mJ ($= 1.6 \times 10^{17}$eV) lasers.
Whether this $ n^3 V \eta $ value is a reasonable assumption or not has to be verified by detailed simulations.
The rate unit for the  Ho figures in the present work
 is $24 \Gamma_0/\epsilon_{eg}^2 = 5.1 \times 10^{-6}  \,{\rm sec}^{-1} \,{\rm eV}^{-2} $, 
assuming $ n^3 V \eta$ of order $ (10^{17})^3\, {\rm cm}^{-6}$ and $\Delta \omega_0/2\pi = 100 {\rm Hz}$.
The very small  rate value  is due to
small combination factors, $\gamma_{pe}^2/\gamma_e $, in particular a small A-coefficient $\gamma_{pe} $.
A possible way to get larger rates is to use multiple identical laser systems of order 10,
each system capable of producing $(10^{17})^3$cm$^{-6}$ of  $ n^3 V \eta$,
within the allowed range of target excitation to $ (10^{20})^3$cm$^{-6}$,
which may enhance the rate unit $24 \Gamma_0/\epsilon_{eg}^2$ to
$5.1 \times 10^{-3}$sec$^{-1}$eV$^{-2}$ due to $\Gamma_0 \propto n^3 V\eta $.
A search for better lanthanoid candidates is highly recommended.

\subsection
{\bf Xe case}

Four lowest excited energy states of Xe are
$^3P_{2,1,0}$ and $^1P_1$ in $LS$ coupling scheme, although 
energy spacings are better described by intermediate coupling scheme close
to $JJ$ scheme.
The following RANP path is used:
\begin{eqnarray}
&&
| e \rangle = 5p^5 (^2P_{1/2}) 6s ^2[1/2]_0 \, 9.4472 {\rm eV} \rightarrow |p\rangle =   
  5p^5 (^2P_{3/2}) 6p ^2[3/2]_1 \,9.7893   {\rm eV} \rightarrow
|q \rangle = | e\rangle
\nonumber \\ &&
\hspace*{1cm}
\rightarrow |g \rangle =  5p^5 (^2P_{3/2}) 6s ^2[3/2]_1 \,8.4365  {\rm eV}
\,,
\end{eqnarray}
where energy values in the free space are taken from NIST data \cite{nist}.
We have in mind using Xe in the free space so that
parity is a good quantum number, and spin-parity $J^P$ changes in Xe RANP path are
$0^- \rightarrow1^+ \rightarrow 0^- \rightarrow 1^-$.
In the proposed scheme of $r=  0.32304 $,
excitation lasers have $\omega_1 = 6.2495\,$ eV $, \omega_2 = 3.1976\,$ eV 
for trigger and   $\omega_0 = \omega =   0.3421 \, $eV for detection.

\begin{figure*}[htbp]
 \begin{center}
 \centerline{\includegraphics{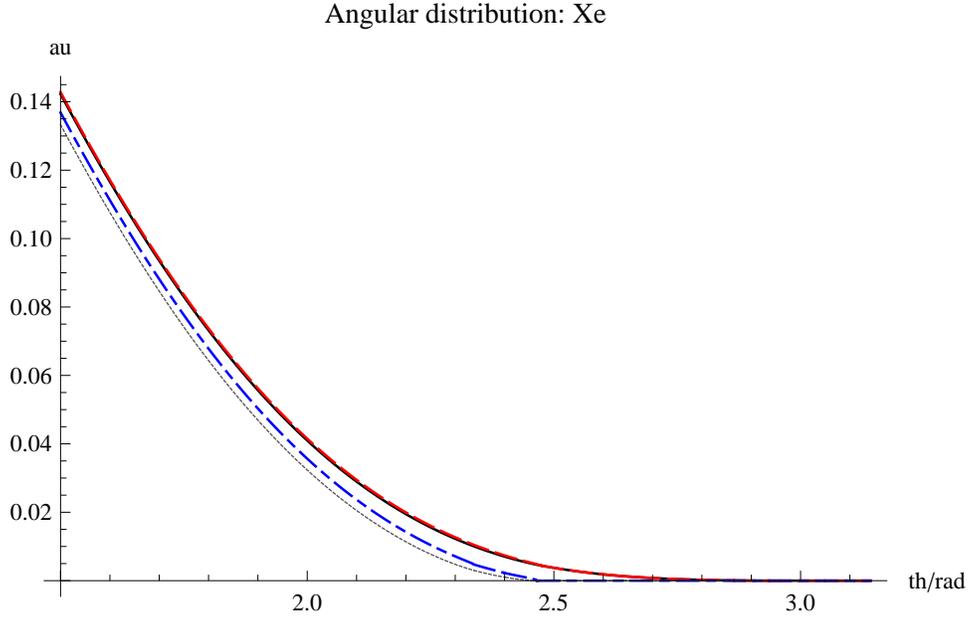}} \hspace*{\fill}
   \caption{ 
Xe RANP angular distribution with $r = 0.32304$.
Cases of NO Majorana of smallest neutrino mass 1 meV in solid black,
NO Majorana of 50 meV in dotted black,
NO Dirac 1 meV in dashed red, and NO Dirac 50 meV in dash-dotted blue.
The absolute rate is obtained by multiplying a factor $0.12\, $sec$^{-1}$ assuming
$n^3 V \eta = (10^{17} )^3 $cm$^{-6}$.
}
   \label {ranp angular distribution xe1}
 \end{center} 
\end{figure*}

\begin{figure*}[htbp]
 \begin{center}
 \centerline{\includegraphics{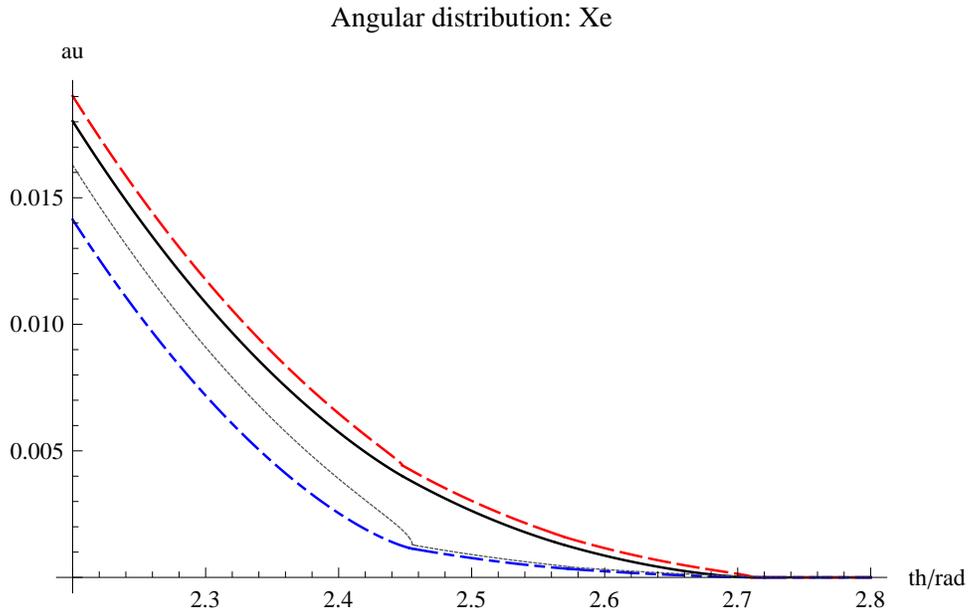}} \hspace*{\fill}
   \caption{ 
Xe RANP angular distribution with $r= 0.32304$.
Cases of NO Majorana of smallest neutrino mass 20 meV in solid black,
NO Dirac of 20 meV in dotted black,
IO Majorana 20 meV in dashed red, and IO Dirac 20 meV in dash-dotted blue.
The absolute rate is obtained by multiplying a factor $0.12\, $sec$^{-1}$ assuming
$n^3 V \eta = (10^{17} )^3 $cm$^{-6}$.
}
   \label {ranp angular distribution nio xe1}
 \end{center} 
\end{figure*}

Xe RANP scale unit is much larger than trivalent Ho ion:
first, the figure of merits factor $F= \gamma_{pe}^2 \epsilon_{eg}^2/ (\gamma_e \epsilon_{pe}^3 ) $ are
\begin{eqnarray}
&&
F({\rm Xe}) = 0.93 \times 10^{-5} 
\,, \hspace{0.5cm}
F({\rm Ho}) = 0.94 \times 10^{-14}
\,.
\end{eqnarray} 
Xe RANP units are
$\Gamma_0 = 5 \times  10^{-3}\,$sec$^{-1}$ and 
$24 \Gamma_0/\epsilon_{eg}^2 = 0.12\,$sec$^{-1}$eV$^{-2}$
using the same value of $n^3V \eta = (10^{17})^3$cm$^{-6}$.
Angular distributions are shown in Fig(\ref{ranp angular distribution xe1}) and Fig(\ref {ranp angular distribution nio xe1}).
Sensitivity to the neutrino mass and Dirac/Majorana distinction is better than the ordinary $2^-$ RENP \cite{renp overview}.

The problem of Xe scheme is a large McQ4 event rate.
Relevant two-photon emission at $|q\rangle \rightarrow | g \rangle $  occurs via E1 $\times $ E1
unlike smaller E1 $\times $ M1 in Ho$^{3+}$ doped crystal.
Xe value of  McQ4 integral is  
 $I_{2\gamma} = 3.1  \times  10^{-22}$eV$^{-2}$ (zero at calculation accuracy for Ho$^{3+}$ case) to
be compared the RANP value,
$I_{2\nu} = 2.3  \times  10^{-49}$eV$^{-2}$.
A solution is to use  photonic crystal for suppression of
strayed McQ4 event \cite{photonic crystal}.
Due to a large level spacing the excitation to Xe $0^-$ is more complicated
than a simple two-photon excitation, which has to be studied.

\vspace{0.5cm}
In summary,
we proposed a general scheme of macro-coherent neutrino-pair emission
stimulated by Raman scattering, 
in order  to measure important neutrino properties, the unknown
smallest neutrino mass to the level of less than 1 meV, NO/IO distinction, and Majorana/Dirac distinction.
The general scheme was illustrated using 
Ho$^{3+}$  doped crystal and Xe atom.
Xe has a larger rate than lanthanoid ions, while
its QED background is much more severe.
Both theoretical and experimental works on QED
background rejection are needed to
make the general scheme realistic. 
It would be interesting to extend the RANP scheme also for search
of other elusive particles such as axion and hidden photon.

\vspace{0.5cm}
 {\bf Acknowledgements}

We thank S. Uetake at Okayama, and C. Braggio,  G. Carugno, and F. Chiossi 
at Padova for useful discussions.
This research was partially
 supported by Grant-in-Aid 17K14363(HH) and  17H02895(MY) from the
 Ministry of Education, Culture, Sports, Science, and Technology.


\begin{thebibliography}{99}
\bibitem{backgrounds}
We assume that major McQn $(n=5, 6)$ events of smaller rates can be
identified by extra photons and are  subtracted from data for RANP,
while McQn ($n\geq 7$)'s  are negligible with very small rates.
Needless to say, higher order spontaneous emission rates without
macro-coherence amplification are completely negligible.
\bibitem{nu parameters}
Parameters determined from neutrino oscillation experiments \cite{pdg} and used this work
are squared mixing matrix elements, $ |U_{ei}|^2, i = 1,2,3$ directly derived from
experimental data, and
mass differences, $\delta m^2_{ij} = m_i^2 - m_j^2 $.
The smallest neutrino mass is assumed in each calculation. Three
CP violation phases  are not known in oscillation experiments, and assumed
vanishing, for simplicity, in the present work.
\end{thebibliography}
\end{document}